\documentclass[mathleft
]{an}
\usepackage{graphicx}
\usepackage{times}
\begin{document}

\Pagespan{789}{}
\Yearpublication{2010}%
\Yearsubmission{2010}%
\Month{11}%
\Volume{999}%
\Issue{88}%

\title{Ground-based observations of {\it Kepler} asteroseismic targets\thanks{Based on observations made with the Isaac Newton Telescope and William Herschel Telescope operated by the Isaac Newton Group, with the Nordic Optical Telescope, operated jointly by Denmark, Finland, Iceland,
Norway, and Sweden, with the Italian Telescopio Nazionale Galileo (TNG) operated  by the Fundaci\'on Galileo Galilei of the INAF (Istituto Nazionale di Astrofisica), and with the Mercator telescope, operated by the Flemish Community,  all on the island of La Palma at the Spanish Observatorio del Roque de los Muchachos of the Instituto de Astrof\'{\i}sica de Canarias. Based on observations made with the IAC-80 operated on the  island of Tenerife by the Instituto de Astrof\'{\i}sica de Canarias  at the Spanish Observatorio del Teide. Also based on observations taken at the observatories of Sierra Nevada, San Pedro M\'artir, Vienna, Xinglong, Apache Point, Lulin, Tautenburg, McDonald, Skinakas, Pic du Midi, Mauna Kea, Steward Observatory, Mt. Wilson, Bia\l{}k\'ow Observatory of the Wroc\l{}aw University, Piszk\'estet\H o Mountain Station, and Observatoire de Haute Provence. Based on spectra taken at the Loiano (INAF-OA Bologna), Serra La Nave (INAF - OA Catania) and Asiago (INAF - OA Padova) Observatories. Also based on observations collected at the Centro Astron\'omico Hispano Alem\'an (CAHA) at Calar Alto, operated jointly by the Max-Planck-Institut f\"ur Astronomie and the Instituto de Astrof\'{\i}sica de Andaluc\'{\i}a (CSIC). We acknowledge with thanks the variable star observations from the AAVSO International Database contributed by observers worldwide and used in this research.}}

\author{K. Uytterhoeven\inst{1}\fnmsep\thanks{\email{katrien.uytterhoeven@cea.fr}}, R. Szab\'o\inst{2}, J. Southworth\inst{3}, S. Randall\inst{4}, R. {\O}stensen\inst{5}, J. Molenda-\.{Z}akowicz\inst{6}, M. Marconi\inst{7}, D.W. Kurtz\inst{8},  L. Kiss\inst{2,9}, J. Guti\'errez-Soto\inst{10}, S. Frandsen\inst{11}, P. De Cat\inst{12}, H. Bruntt\inst{13}, M. Briquet\inst{5}, X.B. Zhang\inst{14}, J.H. Telting\inst{15}, M. St\c{e}\'slicki\inst{6}, V. Ripepi\inst{7}, A. Pigulski\inst{6}, M. Papar\'o\inst{2}, R. Oreiro\inst{5}, C. Ngeow \inst{16}, E. Niemczura\inst{6}, J. Nemec\inst{17}, A. Narwid\inst{6}, P. Mathias\inst{18}, S. Mart\'{\i}n-Ru\'{\i}z\inst{10}, H. Lehmann\inst{19}, G. Kopacki\inst{6}, C. Karoff\inst{20,11}, J. Jackiewicz\inst{21}, M. Ireland\inst{9}, D. Huber\inst{9}, A.A. Henden\inst{22}, G. Handler\inst{23}, A. Grigahc\`ene\inst{24}, E.M. Green\inst{25}, R. Garrido\inst{10}, L. Fox Machado\inst{26}, J. Debosscher\inst{5}, O.L. Creevey\inst{27}, G. Catanzaro\inst{28}, Z. Bogn\'ar\inst{2}, K. Biazzo\inst{29}, S. Bernabei\inst{30}
}

\titlerunning{Ground-based follow-up of {\it Kepler}}
\authorrunning{K. Uytterhoeven et al.}
\institute{
Lab. AIM, CEA/DSM-CNRS-Universit\'e Paris Diderot; CEA, IRFU, SAp, Saclay, 91191, Gif-sur-Yvette, France
\and 
Konkoly Observatory of the Hungarian Academy of Sciences, 1121 Budapest, Hungary
\and 
Department of Physics, University of Warwick, Coventry CV4 7AL, UK
\and
European Southern Observatory, Karl-Schwarzschild-Str. 2, 85748 Garching bei München, Germany
\and 
Instituut voor Sterrenkunde, KULeuven, Celestijnenlaan 200D, 3001 Leuven, Belgium
\and
Instytut Astronomiczny, Uniwersytet Wroc\l{}awski, Kopernika 11, 51-622
Wroc\l{}aw, Poland
\and
INAF - Osservatorio Astronomico di Capodimonte, Via Moiariello 16, 80131 Napoli, Italy
\and
Jeremiah Horrocks Institute of Astrophysics, University of Central Lancashire, Preston PR1 2HE, UK
\and
Sydney Institute for Astrophysics, School of Physics, University of Sydney, Australia
\and
Instituto de Astrof\'{\i}sica de Andaluc\'{\i}a (CSIC), Apartado 3004, 18080 Granada, Spain
\and
Department of Physics and Astronomy, Aarhus University, 8000 Aarhus C, Denmark
\and
Royal Observatory of Belgium, Ringlaan 3, 1180 Brussel, Belgium
\and
LESIA, Observatoire de Paris-Meudon, 92195 Meudon, France
\and
National Astronomical Observatories, Chinese Academy of Sciences, Beijing 100012, China
\and
Nordic Optical Telescope, Santa Cruz de La Palma, Spain
\and
National Central University, No. 300, Jhongda Rd, Jhongli City, Taoyuan County 32001, Taiwan
\and
Department of Physics \& Astronomy, Camosun College, Victoria, British Columbia, Canada
\and
Lab. d'Astrophysique de Toulouse-Tarbes, Universit\'e de Toulouse, CNRS, 57 avenue d'Azereix, 65000 Tarbes, France
\and 
Th\"uringer Landessternwarte, 07778 Tautenburg, Germany
\and
School of Physics and Astronomy, University of Birmingham, Edgbaston, Birmingham B15 2TT, UK
\and
Department of Astronomy, New Mexico State University, Las Cruces, NM 88001, USA
\and
American Association of Variable Star Observers, 49 Bay State Road, Cambridge, MA 02138, USA
\and
Institut f\"ur Astronomie, T\"urkenschanzstr. 17, 1180 Wien, Austria
\and
Centro de Astrof\'{\i}sica, Faculdade de Ci\^encias, Universidade do 
Porto, Rua das Estrelas, 4150-762 Porto, Portugal
\and
Steward Observatory, University of Arizona, 933 North Cherry Avenue, Tucson, AZ 85721, USA
\and
Observatorio Astron\'omico Nacional, Instituto de Astronom\'{\i}a,  UNAM, Ensenada B.C., Apdo. Postal 877, M\'exico 
\and
Instituto de Astrof\'isica de Canarias, 38200 La Laguna, Tenerife, Spain;
Departamento de Astrof\'isica, Universidad de La Laguna, 38205 La
Laguna, Tenerife, Spain
\and
INAF - Osservatorio Astrofisico di Catania, Via S. Sofia 78, 95123 Catania, Italy
\and
INAF - Osservatorio Astrofisico di Arcetri, Largo Enrico Fermi 5, 50125 Firenze, Italy
\and
INAF - Osservatorio Astronomico di Bologna, Via Ranzani
1, 40127 Bologna, Italy
}

\received{01 April 2010}
\accepted{---}
\publonline{later}


\abstract{We present the ground-based activities within the different working groups of the {\it Kepler} Asteroseismic Science Consortium (KASC). The activities aim at the systematic characterization of  the 5000+ KASC targets and at the collection of ground-based follow-up time-series data of selected promising {\it Kepler} pulsators. So far, 36 different instruments at 31 telescopes on 23 different observatories in 12 countries are in use and a total of more than 530 observing nights has been awarded.}

\maketitle

\section{Introduction}

The {\it Kepler} Asteroseismic Science Consortium, KASC\footnote{http://astro.phys.au.dk/KASC}, u\-ni\-tes hundreds of asteroseismologists from institutes all over the world in different topical Working Groups, with the aim of performing seismic studies of all types of pulsating stars across the Hertzsprung-Russell diagram, based on {\it Kepler} time-series space photometry.  The ground-based observational Working Groups (GBOsWG)  take care of the organisation of  ground-based observations in support of the {\it Kepler} space data. Additional ground-based multi-colour and spectral information are indispensable for a successful seismic modelling (see, e.g., Uytterhoeven et al. 2008a, 2009; Uytterhoeven 2009). The need for ground-based support data is motivated by two objectives: 1) the characterization of all {\it Kepler} targets in terms of fundamental stellar parameters, 2) the identification of mode parameters from  multi-colour and spectral time-series observations for selected pulsators. 

The KASC GBOsWG is making great efforts in organising and planning telescope time on various instruments around the world to meet these objectives and to ensure an optimal seismic exploitation of the {\it Kepler} data.  So far, 36 different instruments at 31 telescopes on 23 different observatories in 12 countries are involved and a total of more than 530 observing nights has been awarded. 

\section{Characterization of 5000+ KASC targets}
The {\it Kepler} space data do not provide information on basic stellar parameters such as effective temperature ($T_{\rm eff}$), gravity ($\log g$), metallicity, and the projected rotational velocity ($v \sin i$), which are important to classify the targets and are crucial for successful asteroseismic modelling.  Hence, spectral and multi-colour information are needed to complement the space data. A first effort to compile a catalogue of stellar parameters, derived from Sloan photometry, has been undertaken in the form of the  {\it Kepler} Input Catalogue (KIC, Latham et al. 2005). However, the accuracy of  values of T$_{\rm eff}$ and $\log$g in KIC is generally too low for seismic modelling. Hence, additional ground-based efforts are required. The aim of the KASC GBOsWG is to obtain for each of the 5000+ KASC asteroseismic targets a spectrum with a sufficient resolution to derive $T_{\rm eff}$, $\log g$, micro-turbulence, $v \sin i$ and metallicity (Sousa et al. 2008; Frasca et al. 2006; Bruntt 2009; Niemczura et al. 2009), and multi-colour information to derive reddening, metallicity, and absolute magnitude (Rogers 1995; Kupka \& Bruntt 2001).

The systematic characterization of 5000+ targets requires  a huge observational effort and involves a long-term project, spread out over several instruments. So far, within the KASC GBOsWG, more than 278  nights have been awarded for the characterization project with 26 different instruments on 17 observatories. More  time has been and will be applied for.

\begin{table*}
\begin{footnotesize}
 \begin{center}
\caption{Overview of the awarded observing time for target characterization. Information is given on the observatory, the telescope and instrument, the number of awarded nights (N) or hours (h), the type of targets, and the principal investigator (P.I.) of the proposal. Proposals aimed at the characterization of several pulsators ($\gamma$\,Dor, $\delta$\,Sct, $\beta$\,Cep, Be, solar-like, roAp, and Slowly Pulsating B (SPB) stars,  and stars in clusters) are labelled ``combined''. Spectra that were obtained through a filler programme at the beginning or the end of the night, are indicated as ``filler''. }
\label{characterization}
\begin{tabular}{llrll}\hline
Observatory & Telescope & N  & Target & P.I.\\ \hline
Sierra Nevada (E) & 0.90m photometer & 8N & combined &SM-R\\
San Pedro Martir (MX) & 1.5m photometer & 5N & combined & LFM\\
& 2.12m spectrograph & 2N & combined & LFM\\
Teide (E) & IAC80 CAMELOT & 14N & combined & KU\\
Piszk\'estet\H o (H) & 1.0mRRC CCD & 7N & combined & MP/ZB\\
Calar Alto (E) & 2.2m BUSCA & 5N & combined & KU\\
La Palma (E) & INT WFC & 5N & combined &KU\\
 & NOT FIES & 3N & combined & KU\\
 & Mercator HERMES & 7N & combined & MB\\
Loiano (I) & 1.52m BFOSC & 4N & combined & VR\\
Catania (I) & 0.9m FRESCO & 7N & combined & VR\\
McDonald (USA) & 2.7m cs23& 8N & combined & PDC\\ 
Tautenburg (D) & 2m Coud\'e & 14N & combined & HL \\\hline
Sierra Nevada (E) & 0.9m photometer & 2N & Be stars & JG-S \\
 & 1.52m ALBIREO & 2+10N & Be stars & JG-S \\
Skinakas (GR) & 1.3m spectrograph & 4N & Be stars & JG-S \\
La Palma (E) & NOT Alfosc & 1N & Be stars & JG-S \\
\hline
Catania (I) & 0.9m FRESCO & 3N & $\delta$\,Sct stars& GC\\
Loiano (I) & 1.52m BFOSC & 3N & $\delta$\,Sct stars& VR\\
& & 10N & $\delta$\,Sct stars& GC \\
Asiago (I) & 1.82m AFOSC & 3N & $\delta$\,Sct stars& VR \\
La Palma (E) & TNG SARG & 2h & $\delta$\,Sct stars& VR \\ \hline
Catania (I) & 0.9m FRESCO & 15+15+25+12+25N & solar-like stars & JM-\.{Z}\\
 & 0.9m CCD & 10N & solar-like stars & JM-\.{Z}\\
La Palma (E) & TNG SARG &  12N& solar-like stars& GC \\
& NOT FIES & 2+1.5N & solar-like stars & CK \\
Mauna Kea (USA) & CFHT ESPaDOnS & 10h & solar-like stars & HB \\ 
Pic du Midi (F) & TBL NARVAL & 20h+20h & solar-like stars & HB \\ 
Mt Wilson (USA) & CHARA PAVO & $>$3N & solar-like stars & DH, MI \\
\hline
Steward (USA) & BOK B\&C spectrograph & 10N & compact stars & EMG \\ 
La Palma (E) & WHT ISIS & 4.5N & compact stars & R{\O} \\
& INT IDS & 5+4N & compact stars & RO \\
& NOT FIES & filler & compact stars & JHT \\ \hline
La Palma (E) & NOT FIES & 6N+7N & K giants, roAp stars & SF\\ \hline
Mauna Kea (USA) & CFHT ESPaDOnS & 2h & giants in NGC\,6811 & HB \\
\hline
La Palma (E) & Mercator HERMES & $\sim$45h & binaries with pulsating components & JD \\ \hline
Tautenburg (D) & 2m Coud\'e & filler & SPB, $\beta$\,Cep stars & HL \\ \hline
Haute Provence (F) & 1.92m SOPHIE& filler & $\gamma$\,Dor stars & PM \\ \hline
\end{tabular}
\end{center}
\end{footnotesize}
\end{table*}

\begin{table*}
\begin{footnotesize}
 \begin{center}
\caption{Overview of the awarded time for the collection of multi-colour or spectral time-series of selected promising asteroseismic {\it Kepler} targets. Information is given on the observatory, the telescope and instrument, the number of awarded nights (N), the type of targets, and the principal investigator (P.I.) of the proposal. }
\label{time-series}
\begin{tabular}{llrll}\hline
Observatory & Telescope & N  & Targets & P.I.\\ \hline 
Sierra Nevada (E) & 1.5m CCD & 15N & NGC\,6866 & RG\\
Vienna (A) & 0.8m CCD & 14N & NGC\,6866 & GH\\
Piszk\'estet\H o (H) & 0.9m CCD & 14N & NGC\,6866 & RS\\
Xinglong (CN) & 0.85m CCD & 14N & NGC\,6866 & XZ\\
Bia\l{}k\'ow (PL) & 0.6m CCD & 8+14N & NGC 6866 &  JM-\.{Z} \\
Catania (I) & 0.9m CCD & 8N & NGC 6866 & KB\\ \hline
Sierra Nevada (E) & 1.5m CCD & 15N  & NGC 6811 & RG\\
Vienna (A) & 0.8m CCD &  14N &   NGC 6811 & GH\\
Piszk\'estet\H o (H) & 0.9m CCD & 14N & NGC 6811 & RS\\
Xinglong (CN) & 0.85m CCD & 14N  & NGC 6811 & XZ\\
Bia\l{}k\'ow (PL) & 0.6m CCD &  10N  & NGC 6811 & JM-\.Z\\
Loiano (I) & 1.52m CCD &  10N  & NGC 6811 & HB\\
Catania (I) & 0.9m CCD & 10N & NGC 6811 & JM-\.Z\\
Teide (E) & IAC-80 CAMELOT &  14N  & NGC 6811 & OC\\
Apache Point (USA) & NMSU 1.0m &  14N  & NGC 6811 & JJ\\  \hline
Lulin (TW) & 0.4m SLT & 18N & RR Lyr, Cepheids & NCC \\
Lulin (TW) & 1.0m LOT & 3N & RR Lyr, Cepheids & NCC \\ 
AAVSONet & 0.2-0.6m telescopes & $>$1N & RR Lyr, Cepheids & AH \\ \hline
Sierra Nevada (E) & 0.9m photometer & 14N & hybrid $\gamma$\,Dor/$\delta$\,Sct  stars& AG/SM-R \\ \hline
McDonald (USA) & 2.2m B\&C spectrograph & 7N & SPB, $\gamma$\,Dor stars& PDC\\ \hline
La Palma (E) & Mercator HERMES & 11N & SPB, $\beta$\,Cep stars & HL \\ \hline
\end{tabular}
\end{center}
\end{footnotesize}
\end{table*}

The first effort to characterize asteroseismic {\it Kepler} targets dates back to 2004. Since then, a project is running to characterize KASC solar-like stars (Mo\-len\-da-\.{Z}a\-ko\-wicz et al. 2007, 2008, 2009b). Nowadays, several observational projects, focussed either on a specific pulsation class or on several classes simultaneously, are ongoing to systematically observe all KASC targets.
In Table~\ref{characterization} we present an overview of the {\it awarded} observing time for target characterization. Additional information on the observations is given in Uytterhoeven et al. (2010). In addition to the spectroscopic and multi-colour observations, an interferometric project is ongoing with PAVO@CHARA at Mt Wilson Observatory (USA) to measure angular diameters for some of the brighest Kepler targets.  Results on the physical parameter determination of a selection of $\delta$\,Sct, $\gamma$\,Dor and hybrid targets are recently presented in Catanzaro et al. (2010).

More observing time has been applied for. Spectropolarimetric observations are planned to investigate magnetic signatures  in selected  Cepheids, RR Lyr, $\delta$\,Sct, and Be stars with ESPaDOnS@CFHT, Mauna Kea (USA) (P.I. JN, JG-S). An ambitious proposal to observe 95\% of all KASC asteroseismic targets with the multi-fiber, multi-object spectrograph LAMOST@4m telescope at Xinglong observatory (CN)  has been submitted (P.I. PDC). 

\section{Time-series observations of selected promising {\it Kepler} pulsators}
Important key ingredients for an asteroseismic study are precise pulsation frequencies, accurately identified pulsation modes, and strong constraints on   atmospheric parameters. Accurate values of the pulsation frequencies will be provided for by the {\it Kepler} photometry, while accurate atmospheric parameters will be derived from the ground-based data obtained in the framework of the project outlined in the previous section. 

For solar-like oscillators, mode identification relies on the regularity of the frequency pattern in the power spectrum (e.g. Mathur et al. 2010). This method is not directly applicable to larger amplitude pulsators, for which a combination of non-linear effects, rotation, and convection selects the observed modes in a way that is not yet fully understood (e.g. Townsend 2009; Miglio et al. 2008; Su\'arez et al. 2005; Degroote et al. 2010). For these targets, the identification of modes observed by {\it Kepler} requires ground-based multi-colour and spectral time-series analysis (e.g. Briquet et al. 2009; Poretti et al. 2009; Uytterhoeven et al. 2008b; Rodr\'{\i}guez et al. 2006). 

Multi-epoch spectroscopy is also important in the case of (eclipsing) spectroscopic binaries with a pulsating component, because by using spectra one can directly derive the component masses (Tango et al. 2006;  Vu\v{c}kovic et al. 2007; Creevey et al. 2009; Desmet et al. 2010), and it is possible to disentangle the binary components (Harmanec et al. 2004) and study the line-profile variability of the components in full detail (Uytterhoeven et al. 2005).

 To date, within the KASC GBOsWG, a total of at least 256 nights has been awarded with 15 different instruments on 13 observatories for specific time-series projects. Additional telescope time has been applied for. An overview of the {\it awarded} observing time is given in Table~\ref{time-series}. We refer again to Uytterhoeven et al. (2010) for a description of the observations. The projects involve RR Lyr stars and Cepheids, Slowly Pulsating B stars, $\beta$\, Cep stars, hybrid $\gamma$\,Dor/$\delta$\,Sct candidates, and pulsators in clusters. The latter concerns a large  photometric multi-site campaign on the clusters NGC\,6866, carried out in 2009,  and NGC 6811, scheduled for 2010. The cluster NGC\,6866  is known to host at least three $\delta$\,Sct and two $\gamma$\,Dor candidates (Mo\-le\-nda-\.{Z}a\-ko\-wicz et al. 2009a), and there are 12 known $\delta$\,Sct stars in NGC\,6811 (Luo et al. 2009).

\section{Future plans}
 The ground-based counterpart of {\it Kepler} is crucial for the successful execution of seismic studies. The GBOsWG will continue to organise ground-based observations to complement the {\it Kepler} light curves. So far, the observational and organisational efforts have been very successful with more than 530 observing nights already awarded. Additional observing time with dedicated multi-colour and spectroscopic instruments will be applied for in the coming observing se\-mes\-ters. 
The ground-based support of {\it Kepler} is putting a heavy pressure on ground-based telescopes in the Northern hemisphere, especially on the ones equipped with a (high-R) spectrograph. Therefore,  assistance and help from the community is very welcome. We encourage everyone who has access to (further) telescopes and wants to help with observations, data reduction or data analysis, to join the project. This very important task of supporting {\it Kepler} from the ground revives the use of small/mid-sized telescopes, which is a significant benefit for all the national observatories involved. 

\begin{acknowledgements}
 MB is Postdoctoral Fellow of the Fund for Scientific Research, Flanders. This work was supported by  MNiSW grant N203 014 31/2650 and by the National Office for Research and Technology through the Hungarian Space Office Grant No. URK09350 and the `Lend\"ulet' program of the Hungarian Academy of Sciences.
\end{acknowledgements}

\end{document}